\documentclass[groupedaddress,fleqn]{article}
\usepackage[margin=1in]{geometry}
\usepackage{graphicx}
\usepackage{amsmath}
\usepackage{multicol}
\usepackage[T1]{fontenc}
\usepackage{ae,aecompl}
\usepackage{color}
\usepackage{authblk}
\usepackage{cite}
\usepackage{caption}

\begin{document}
	\title{Nonlinear variations in spherically symmetric accretion in the Schwarzschild metric }
	\author{Md Arif Shaikh\thanks{arifshaikh@hri.res.in}}%
	\affil{\sl \small Harish-Chandra Research Institute, HBNI, Chhatnag Road, Jhunsi, Allahabad 211019, India}
	\date{\today}
	\maketitle
	
\begin{abstract}
	\noindent In this work, we study the implications of nonlinearity in general relativistic spherically symmetric inviscid irrotational accretion flow in a stationary non-rotating spacetime. It has been found that the perturbation scheme leads to a differential equation of the form of general Li{\'e}nard's equation. We discuss the equilibrium conditions of this system and its implications for globally subsonic accretion flows in the spherically symmetric stationary background. It is found that the stable solution predicted by linear stability analysis may become unstable under inclusion of lowest order nonlinearity.
\end{abstract}	

\begin{multicols}{2}

\section{Introduction}
The spherically symmetric stationary inviscid hydrodynamic accretion flow on to an accretor in a background Newtonian potential is known as the Bondi flow \cite{Bondi1952}. The basic equations to describe such a flow is the mass conservation equation (the continuity equation ) and the momentum conservation equation (the Euler equation). These are partial differential equations of both time and position and are nonlinear in nature. The first attempt to simplify these equations is to consider a case where the flow variables remain effectively same over the period of astrophysical observation. By flow variables, for example, we mean the fluid bulk velocity, the sound speed, the density of the fluid or other astrophysically important variables which are functions of the velocity and density. Such flow with effectively no time dependence is called stationary flow. The Bondi flow is basically such a stationary accretion flow. For stationary accretion flow, the two governing equations, upon integration, give two conserved quantities. Continuity equation gives the mass accretion rate which is the rate of infall of matter across the surface of a spherical shell. The momentum conservation equation gives the Bernoulli's constant which for flow governed by the adiabatic equation of state is the specific energy of the fluid. 

The Bondi flow as mentioned earlier is a stationary solution of the partial differential equations describing hydrodynamics of the infalling matter. Thus before using such solution for the practical purpose, one has to make sure that such solution is stable. The most basic task one performs to check the stability of a particular nonlinear system is called linear stability analysis. In linear stability analysis, one introduces a small perturbation which is a function of time and position, to the accretion variables. In other words, instead of taking the accretion variables to be time-independent, one writes the accretion variables as a sum of a time-independent part and a time-dependent part. The accretion variables in the governing equations are substituted by this sum of the stationary part and the small time-dependent part and at all stages, the terms that are higher than the linear order in the perturbations are neglected. In other words, all the equations are linear in the perturbations. The resulting equations could be manipulated to have a wave equation of the perturbation.
One can perturb different accretion flow variables to obtain such wave equation of linear perturbation. The most common linear perturbation scheme is to obtain the wave equation of perturbation of the velocity potential field of the irrotational flow\cite{Moncrief1980,Petterson1980,Unruh,Visser1998,Bilic1999}. 
however, one can also obtain such wave equation of perturbation of the mass accretion rate \cite{Deepika2015,Deepika2017,Shaikh2017,Shaikh2018a} or the Bernoulli's constant\cite{Shaikh2017,Shaikh2018a,Datta2018}.

It was Moncrief \cite{Moncrief1980} who first showed that the linear perturbation of potential field of an irrotational flow leads to a wave equation which mimics the wave equation of a scalar field in curved spacetime. Thus, the propagation of the acoustic perturbation is governed by an acoustic spacetime metric which is curved. Later it was shown by Unruh \cite{Unruh} that for transonic flow, the acoustic metric possesses an acoustic horizon and there is an analogous Hawking radiation from the sonic horizon. This opened up a new field of research which is known as `Analogue gravity phenomena' \cite{Unruh,Visser1998,Bilic1999,Barcelo,Novello-visser,Unruh-Schutzhold,Analogue-gravity-phenomenology}. However, one crucial point is that the emergence of analogue gravity phenomena depends on the scheme of the perturbation analysis, i.e., it comes as an outcome of {\it linear} stability analysis. Therefore, it is to be seen whether such phenomena arise even in the case of nonlinear perturbation analysis.

The linear perturbation analysis, though may be the first step to check the stability of stationary accretion flow, it is not totally reliable. This is due to the fact that common knowledge from nonlinear system tells us that though a nonlinear system is stable under linear perturbation, it may lose the stability under influence of next order nonlinearity. Though there have been numerical simulations to study such effects, analytical work on this topic is still very rare. Recently, Sen and Ray \cite{Sen2014} studied the effects of nonlinearity for spherically symmetric Newtonian accretion by introducing arbitrary order nonlinearity while constructing the wave equation of the perturbation of mass accretion rate. In \cite{Sen2014}, Sen and Ray study the stability of Bondi flow which, as mentioned earlier, is the spherically symmetric inviscid Newtonian accretion flow. In such system, the influence of gravity is incorporated by prescribing a gravitational potential field. Such description of accretion flow is not sufficient in the vicinity of strong gravity where the spacetime deviates from the flat Minkowski space and becomes curved. To describe such accretion flow one has to use the full general relativistic approach. The general relativistic version of Bondi flow, i.e., the general relativistic spherically symmetric inviscid hydrodynamic accretion in a spherically symmetric stationary background spacetime was given by Michel\cite{Michel1972} and is often referred to as the Michel flow.

In the present work, we study the stability of the Michel flow using a perturbation scheme similar to that in \cite{Sen2014}. It is noticed that the wave equation (See equation (\ref{wave-eqn})) is similar in form to that found in the linear stability analysis. However, the metric elements $ F^{\mu\nu} $ (defined in equation (\ref{F}) ) contains the full accretion variables and not only the stationary part as in case of linear stability analysis. Also $ F^{\mu\nu} $ is still symmetric as in linear stability analysis. Thus in this perturbation scheme, $ F^{\mu\nu} $ and hence the wave equation contains nonlinearity of arbitrary order. To find an expression fully in terms of the perturbation of mass accretion rate the nonlinearity is kept only up to the lowest order. Thus such equation while containing the lowest order nonlinearity becomes easy to handle as well as gives us a glimpse of the implications that nonlinearity has for such accretion flow.

In Section \ref{Sec:gov-eq}, we provide the basic equations needed to describe the general relativistic accretion flow and define relevant thermodynamic quantities. In section \ref{Sec:nonlin-pert}, we perform the perturbation analysis containing an arbitrary order of nonlinearity and find the expression for the wave equation. In section \ref{Sec:standing-wave-analysis}, we use the wave equation found in section \ref{Sec:nonlin-pert} to obtain an equation fully in terms of the perturbation of mass accretion rate and use this wave equation to study globally subsonic flows which leads to the equation of the form of general Li{\'e}nard's equation.

We shall set $ G = c = M = 1 $, where $ G $ is the universal gravitational constant, $ c $ is the velocity of light and $ M $ is the mass of the accretor. Radial distance is scaled by $ GM/c^2 $ and velocities are scaled by $ c $. We shall use negative-time-positive-space metric convention.
\section{Governing Equations}\label{Sec:gov-eq}
We consider spherically symmetric metric given by
\begin{equation}\label{metric}
ds^2 = -g_{tt}dt^2+g_{rr}dr^2+g_{\theta\theta}d\theta^2+g_{\phi\phi}d\phi^2
\end{equation}
where the metric elements are given by
\begin{equation}\label{key}
g_{tt}=g_{rr}^{-1}=(1-2/r),\quad g_{\theta\theta} = g_{\phi\phi}/\sin^2\theta = r^2
\end{equation}
The fluid is assumed to be perfect and the energy momentum tensor is given by
\begin{equation}\label{eng-mom-ten}
T^{\mu\nu} = (p+\varepsilon)v^\mu v^\nu + p g^{\mu\nu}
\end{equation}
where $ p $ is the pressure and $ \varepsilon $ is the energy density of the fluid which consists of the rest-mass energy density plus the thermal energy density. Pressure and the density $ \rho $ is related by the equation of state, for adiabatic fluid which is given by the relation $ p = K \rho^\gamma $. $ K $ is a constant for isentropic fluid and $ \gamma $ is the ratio of the specific heat at constant pressure ($ c_p $) and that at constant volume $ (c_v) $. $ v^\mu $ is the four-velocity of the fluid and obeys the normalization condition $ v_\mu v^\mu = -1 $. The accretion flow is governed by the continuity equation
\begin{equation}\label{cont}
 \nabla_\mu (\rho v^\mu) = 0
\end{equation}
and the energy momentum conservation equation
\begin{equation}\label{eng-mom}
\nabla_\mu T^{\mu\nu} = 0
\end{equation}
The thermodynamic enthalpy is given by
  \begin{equation}\label{enthalpy}
  h = \frac{p+\varepsilon}{\rho}
  \end{equation}
and the sound speed is defined as
\begin{equation}\label{cs}
	c_{s}^2 = \left.\frac{\partial p}{\partial \varepsilon}\right|_{\rm constant~entropy} = \frac{\rho}{h}\frac{\partial h}{\partial \rho}
\end{equation}
The irrotationality condition gives \cite{Bilic1999}
\begin{equation}\label{irrotational}
	\partial_\mu(hv_\nu)-\partial_\nu (hv_\mu) = 0
\end{equation}
\section{Nonlinearity in  perturbation analysis}\label{Sec:nonlin-pert}
For spherically symmetric accretion, $ \partial_\theta= 0 =  \partial_\phi$, hence the continuity equation (\ref{cont}) becomes
\begin{equation}\label{cont-sph}
	\partial_t (\sqrt{-g} \rho v^t )+\partial_r(\sqrt{-g}\rho v^r) = 0
\end{equation}
For stationary flow ($ \partial_t = 0 $), one has $ f_0 = \sqrt{-g}\rho_0 v^r_0 = {\rm constant} $. $ f_0 $ is basically the stationary mass accretion rate divided by $ 4\pi $ ( the geometrical factor arising due to integral over $ \theta $ and $ \phi $). To perform the perturbation analysis we write each variables as the sum of two parts, one is the stationary part (time independent) and another is the time dependent part. Thus we write
\begin{equation}\label{variables}
\begin{aligned}
& v^t = v^t_0(r)+(v^t)'(r,t)\\
& v^r = v^r_0(r)+(v^r)'(r,t)\\
& \rho = \rho_0(r)+\rho'(r,t)
\end{aligned}
\end{equation}
We now define a variable as $ f(r,t) = \sqrt{-g}\rho v^r $ which could be written as
\begin{equation}\label{f}
f(r,t)=f_0 + f'(r,t)
\end{equation}
where $ f' $ can be written as
\begin{equation}\label{f'}
\frac{f'}{f_0} = \frac{\rho'}{\rho_0}+\frac{(v^r)'}{v^r_0}+\frac{\rho' (v^r)'}{\rho_0 v^r_0}
\end{equation}
Using the variables as defined in equation (\ref{variables}) in the continuity equation (\ref{cont-sph}) and collecting the terms gives
\begin{equation}\label{delrf}
v^t\partial_t\rho' + \rho \partial_t (v^t)' = - \frac{1}{\sqrt{-g}}\partial_r f'
\end{equation}
differentiating the normalization condition $ v_\mu v^\mu = -1 $ with respect to $ t $ gives $ \partial_t(v^t)' $ in terms of $ \partial_t(v^r)' $ as $ \partial_t(v^t)' = \alpha(r) \partial_t (v^r)'$, where $ \alpha(r) = (g_{rr}v^r)/(g_{tt}v^t)  $. Thus equation (\ref{delrf}) can be rewritten as
\begin{equation}\label{delrf2}
v^t\partial_t\rho' + \alpha\rho \partial_t (v^r)' = - \frac{1}{\sqrt{-g}}\partial_r f'
\end{equation}
also differentiating equation (\ref{f'}) with respect to $ t $ gives
\begin{equation}\label{deltf}
v^r\partial_t \rho' + \rho \partial_t (v^r)' = \frac{1}{\sqrt{-g}}\partial_t f'
\end{equation}
From equation (\ref{delrf2}) and (\ref{deltf}) we find
\begin{equation}\label{deltrhoanddeltv}
\begin{aligned}
& \partial_t\rho' = -\frac{1}{\sqrt{-g}}\left[g_{rr}v^r \partial_t f' + g_{tt}v^t\partial_r f'\right]\\
& \partial_t(v^r)' = \frac{g_{tt}v^t}{\sqrt{-g}\rho}\left[v^t\partial_t f'+v^r \partial_r f'\right]
\end{aligned}
\end{equation}

We set $ \mu = t $, $ \nu = r $ in equation (\ref{irrotational}) and divide by $ hv_t $. Differentiating the resulting equation with respect to $ t $ and using $ \partial_t v^r = \partial_t(v^r)'$ and $ \partial_t\rho = \partial_t \rho' $ gives
\begin{equation}
\begin{aligned}
& \partial_t\left[\frac{g_{rr}}{g_{tt}v^t}\partial_t (v^r)'\right] + \partial_t \left[\frac{g_{rr}v^r c_{s}^2}{g_{tt}\rho v^t}\partial_t \rho'\right]\\
& + \partial_r\left[\frac{g_{rr}v^r}{g_{tt}(v^t)^2}\partial_t (v^r)'\right] + \partial_r\left[\frac{c_s^2}{\rho}\partial_t\rho'\right]=0
\end{aligned}
\end{equation}
Finally using equations in (\ref{deltrhoanddeltv}) in the above equation gives 
\begin{equation}\label{wave-eqn}
\partial_\mu (F^{\mu\nu}\partial_\nu f') = 0
\end{equation}
where $ F^{\mu\nu} $ is given by
\begin{equation}\label{F}
\begin{aligned}
& F^{\mu\nu} = \frac{g_{rr}v^r}{fv^t} \times \\
& \begin{bmatrix}
{c_s^2g^{tt}+(1-c_s^2)(v^t)^2} & v^rv^t(1-c_s^2)\\
v^rv^t(1-c_s^2) & {-c_s^2g^{rr}+(1-c_s^2)(v^r)^2}
\end{bmatrix}
\end{aligned}
\end{equation}
where $ \mu,\nu $ run from $ 0 $ to $ 1 $ with $ 0 $ and $ 1 $ standing for $ t $ and $ r $ respectively.
$ F^{\mu\nu} $ contains the full variables and not only the stationary part. The results of linear perturbation is readily achieved by using the stationary parts only in the $ F^{\mu\nu} $ elements\cite{Bilic1999,Deepika2015,Shaikh2017,Datta2018} and may be given by
\begin{equation}
\begin{aligned}\label{linearF}
& F^{\mu\nu}_0 = \frac{g_{rr}v^r_0}{fv_0^t} \times \\
& \begin{bmatrix}
{c_{s0}^2g^{tt}+(1-c_{s0}^2)(v^t_0)^2} & v^r_0v^t_0(1-c_{s0}^2)\\
v^r_0v^t_0(1-c_{s0}^2) & {-c_{s0}^2g^{rr}+(1-c_{s0}^2)(v^r_0)^2}
\end{bmatrix}
\end{aligned}
\end{equation}
 Also similar to the non relativistic Newtonian case, here also $ F^{\mu\nu} $ is symmetric in form. The Newtonian results \cite{Sen2014} are obtained by taking the limit $ g_{rr}=g^{tt}=1/g^{rr} \to 1 $ and $ c_{s}\ll 1 $, $ v^r\ll 1 $ which gives
\begin{equation}\label{key}
F^{tt} = \frac{v^r}{f},\quad F^{tr}=F^{rt}=\frac{(v^r)^2}{f},\quad F^{rr} = \frac{v^r}{f}((v^r)^2-c_s^2)
\end{equation}

In the linear stability analysis, the wave equation (\ref{wave-eqn}) becomes $ \partial_\mu (F^{\mu\nu}_0\partial_\nu f') = 0 $. This equation is similar to the wave equation of a massless scalar field $ \psi $ in curved space time give by 
\begin{equation}
\partial_\mu (\sqrt{-g}g^{\mu\nu}\partial_\nu \psi) = 0
\end{equation}
where $ g_{\mu\nu} $ is the spacetime metric and $ g $ is the determinant of $ g_{\mu\nu} $. Comparing these two equations, one makes an analogy that the propagation of the acoustic perturbation is described by an analogue spacetime metric $ G^{\mu\nu} $ where $ \sqrt{-G}G^{\mu\nu} = F^{\mu\nu} $, $ G $ being the determinant of $ G_{\mu\nu} $. Such emergence of an analogue spacetime metric in linear perturbation fluid is known as analogue gravity phenomena. The time-independent metric $ G_{\mu\nu} $  possesses an acoustic horizon similar to the event horizon of a black hole. The acoustic horizon acts as a way one-way membrane for the acoustic perturbation. It can be showed that the acoustic horizon actually coincides with the transonic surface where the fluid bulk velocity and the sound speed becomes equal. The acoustic horizon separates the subsonic region from supersonic region. The acoustic perturbations cannot come out from the supersonic region to the subsonic region. This is due to the fact that in the supersonic region the bulk velocity of the fluid is greater than the speed of the acoustic perturbation and the acoustic perturbation is basically dragged with the fluid medium. As no acoustic perturbation can come out of the transonic region, it is often term as a `dumb hole' in analogy to the `black hole' from which even light can not escape.

However, this analogy breaks down when we include nonlinearity in the perturbation scheme. This is because the nonlinear terms are time-dependent and the metric $ F^{\mu\nu} $ becomes time-dependent and the above analogy no longer stands. This breaking down of the analogy between black hole event horizon and acoustic horizon has been studied numerically by Mach and Malec \cite{Mach2008}.

\section{Standing wave analysis for gloabally subsonic flows}\label{Sec:standing-wave-analysis}
Equation (\ref{wave-eqn}) gives a nonlinear equation of perturbation with nonlinearity of arbitrary order. The wave equation (\ref{wave-eqn}) already contains a term $ \partial_\mu f' $ which is linear in $ f' $.
Therefore the results of linear perturbation analysis would be obtained if in the $ F^{\mu\nu} $ elements we retain only stationary part of the accretion variables, i.e., $ v^r=v^r_0, c_s = c_{s0} $ and $ \rho=\rho_0 $, as mentioned in the previous section. Similarly, the lowest order of nonlinearity (i.e., terms that are second order in the perturbations) will be introduced in the wave equation (\ref{wave-eqn}) if we keep only the terms that are linear in perturbations in $ F^{\mu\nu} $, i.e., if we neglect terms like $ ((v^r)')^2,(\rho')^2,((v^r)'\rho') $ and higher order terms. In the following, we will perform second order stability analysis by keeping terms that are up to second order in the perturbations. Thus, now equation (\ref{f'}) will become
\begin{equation}\label{linf'}
\frac{f'}{f_0} = \frac{\rho'}{\rho_0}+\frac{(v^r)'}{v^r_0}
\end{equation}
We want to obtain the wave equation (\ref{wave-eqn}) fully in terms of $ f' $ and other stationary variables. Thus we would like to obtain expressions for $ (v^r)' $ and $ \rho' $ in terms of $ f' $.
Now in standing wave analysis, it is common to express the perturbations as multiplicatively separable functions of time and space, with an exponential time part. Using such scheme in equation (\ref{delrf2})
gives the following relation
\begin{equation}\label{linrho'}
\frac{\rho'}{\rho_0} + \sigma_1(r)\frac{(v^r)'}{v^r_0} = \sigma_2(r)\frac{f'}{f_0}
\end{equation}
where $ \sigma_1(r) $ and $ \sigma_2(r) $ depends on spatial part of the perturbations $ v',\rho', $ and $ f' $. From equations (\ref{linf'}) and (\ref{linrho'}) we find
\begin{equation}\label{rho'v'}
\begin{aligned}
& \frac{\rho'}{\rho_0} = \sigma(r)\frac{f'}{f_0},\quad \sigma(r) = \frac{\sigma_2(r)-\sigma_1(r)}{1-\sigma_1(r)}\\
& \frac{(v^r)'}{v^r_0} = \tilde{\sigma}(r)\frac{f'}{f_0},\quad \tilde{\sigma}(r) = \frac{1-\sigma_2(r)}{1-\sigma_1(r)}
\end{aligned}
\end{equation}
Finally using expressions in equation (\ref{rho'v'}), $ F^{\mu\nu} $ elements could be written as
\begin{equation}\label{F2}
\begin{aligned}
& F^{tt} = \frac{g_{rr}v^r_0}{f_0 v^t_0}\left[\left(c_{s0}^2g^{tt}+(1-c_{s0}^2)(v^t_0)^2\right)+\epsilon \xi^{tt}\frac{f'}{f_0}\right]\\
& F^{tr} = \frac{g_{rr}v^r_0}{f_0 v^t_0}\left[(1-c_{s0}^2)v^t_0 v^r_0 +\epsilon \xi^{tr}\frac{f'}{f_0}\right] \\
& F^{rt} = \frac{g_{rr}v^r_0}{f_0 v^t_0}\left[(1-c_{s0}^2)v^r_0v^t_0 +\epsilon \xi^{rt}\frac{f'}{f_0}\right] \\
& F^{rr} = \frac{g_{rr}v^r_0}{f_0 v^t_0}\left[\vphantom{\frac{f'}{f_0}}(-c_{s0}^2g^{rr}+(1-c_{s0}^2)(v^r_0)^2)\right.\\
&\left. +\epsilon  \xi^{rr}\frac{f'}{f_0}\right]
\end{aligned}
\end{equation}
where $ \epsilon $ has been introduced to work as a switch, $ \epsilon = 0 $ gives the result for linear stability analysis and $ \epsilon=1 $ gives the lowest order nonlinear analysis. $ \xi^{\mu\nu} $ are given by
\begin{equation}\label{key}
\begin{aligned}
& \xi^{tt} = \left[\left(\frac{\tilde{\sigma}}{g_{tt}(v^t_0)^2}-1\right)\left(c_{s0}^2g^{tt}+(1-c_{s0}^2)(v^t_0)^2\right)\right.\\
& \left. +c_{s0}^2\left(\beta \sigma(r)(g_{rr}/g_{tt})(v^r_0)^2-2v^t_0v^r_0 \alpha \tilde{\sigma}\right)  \vphantom{\left(\frac{\tilde{\sigma}}{g_{tt}(v^t_0)^2}-1\right)}\right]\\
& \xi^{tr} = \xi^{rt} = v^r_0\left[\left(\frac{\tilde{\sigma}}{g_{tt}(v^t_0)^2}-1\right)v^t_0(1-c_{s0}^2)\right.\\
&+\tilde{\sigma}\left(\frac{-1+2g_{tt}(v^t_0)^2}{g_{tt}v^t_0}\right)
\left. +c_{s0}^2(\beta \sigma v^t_0 - \tilde{\sigma}(v^t_0+\alpha v^r_0))\vphantom{\left(\frac{\tilde{\sigma}}{g_{tt}(v^t_0)^2}-1\right)}\right]\\
& \xi^{rr} = (v^r_0)^2\left[\left\{-g^{rr}\left(\frac{\tilde{\sigma}}{g_{tt}(v^t_0)^2}-1\right)-\frac{g_{tt}(v^t_0)^2}{g_{rr}}\beta \sigma\right. \right. \\ 
& \left.\left.-2\tilde{\sigma}(v^r_0)^2 \vphantom{\left(\frac{\tilde{\sigma}}{g_{tt}(v^t_0)^2}-1\right)} \right\}\frac{c_{s0}^2}{(v^r_0)^2}. + (1-c_{s0}^2)\left(\frac{\tilde{\sigma}}{g_{tt}(v^t_0)^2}-1\right)+2\tilde{\sigma}\right]
\end{aligned}
\end{equation}
where $ \beta $ comes from the perturbation of $ c_{s}^2 $, $ (c_{s}^2)'/c_{s0}^2 = \beta (\rho'/\rho_0)$ and is given by
\begin{equation}\label{beta}
\beta = \frac{\gamma(\gamma-1)(\gamma-1-c_{s0}^2)}{\gamma(\gamma-1-c_{s0}^2)+c_{s0}^2}
\end{equation}
The Newtonian limit can be obtained by taking the limit $ g_{rr}=1/g^{rr}=1/g_{tt}=g^{tt}\to 1, c_{s0}\ll1,v^r_0\ll1, v^t_0\to 1 $ and $ \sigma = \sigma_2,\tilde{\sigma} = 1-\sigma_2 $.

Now we use $ F^{\mu\nu} $ given by equation (\ref{F2}) to get the wave equation (\ref{wave-eqn}) in terms of $ f' $ and the stationary accretion variables. In order to have a simplified form let us write $ F^{\mu\nu}  $ as $ F^{\mu\nu}=F^{\mu\nu}_0 + \epsilon F^{\mu\nu}_1 f'$, where $ F^{\mu\nu} _0$ is the  the matrix $ F^{\mu\nu} $ with the accretion variables having stationary values given by equation (\ref{linearF}) and $ F_1^{\mu\nu} $  comes as we want to introduce nonlinearity in the wave equation. $ F^{\mu\nu}_0 $ and $ F^{\mu\nu}_1 $ can be read from the equation (\ref{F2}). In particular $ F^{\mu\nu}_1 $ is given by
\begin{equation}\label{F1}
F^{\mu\nu}_1 = \frac{g_{rr}v^r_0}{f_0^2 v^t_0}\xi^{\mu\nu}
\end{equation}

In intermediate step, we divide the whole equation by $ F^{tt}_0(1+\epsilon F f') $, where $ F = F^{tt}_1/F^{tt}_0 $, to make the coefficient of $ \ddot{f}' $ equal  to unity. `Dot' stands for partial derivative with respect to $ t $.
As we are interested in retaining only the terms that are up to second order in nonlinearity. Dividing by $ (1+\epsilon F f') $ is equivalent to multiplication by $ (1-\epsilon F f') $. Such manipulation gives the following equation

\begin{equation}
\begin{aligned}
& \ddot{f}' + 2 (F^{tr}_0/F^{tt}_0)\partial_r\dot{f}' + (\partial_r F^{tr}_0/F^{tt}_0)\dot{f}'+(1/F^{tt}_0) \\
& \times \partial_r\left(F^{rr}_0 \partial_r f'\right) + \frac{\epsilon }{F^{tt}_0}\left[-2F F^{tr}_0f'\partial_r\dot{f}'-Ff'\dot{f}'\partial_rF^{tr}_0 \right.\\
& \left. -Ff' \partial_r\left(F^{rr}_0\partial_rf'\right)+F^{tt}_1(\dot{f}')^2+\partial_r\left(F^{tr}_1\partial_t(f')^2\right)\right.\\
&\left. -\frac{1}{2}\partial_r F^{tr}_1 \partial_t(f')^2+\frac{1}{2}\partial_r\left(F^{rr}_1 \partial_r(f')^2\right)\right] = 0
\end{aligned}
\end{equation}

$ \epsilon=0 $ gives the familiar equation discussed in linear stability analysis for general relativistic accretion \cite{Deepika2015,Deepika2017,Shaikh2018a,Datta2018}. Up to this point, the equations are valid for any kind of flow, i.e., it may be subsonic or supersonic. Below, we use it for the globally subsonic flows.

Let us use a trial solution of the form $ f' = R(r)\Phi(t) $. We multiply the whole equation by $ R F^{tt}_0 $ and rearrange the terms. In standing wave, the boundary conditions dictate that the wave amplitude becomes zero at the two boundary points. Thus we try to rewrite the terms as total derivatives with respect to $ r $ because such terms upon integration over the region between the boundary points would appear as surface terms which vanish at the boundary points. This provides the following equation
\begin{equation}\label{Rphi}
\begin{aligned}
& R^2 F^{tt}_0 \ddot{\Phi}+\dot{\Phi}\partial_r\left(R^2 F^{tr}_0\right)+\Phi \left\{\partial_r\left(\frac{1}{2}F^{rr}_0\partial_r R^2\right) \right. \\
& \left. -F^{rr}_0\left(\partial_r R\right)^2 \vphantom{\left(\frac{1}{2}F^{rr}_0\partial_r R^2\right) }\right\} + \epsilon \left[ \dot{\Phi}^2 R^3 F^{tt}_1 + \Phi \dot{\Phi} \left\{ \vphantom{\frac{F^{tr}_1}{3}} \partial_r (R^3 F^{tr}_1) \right. \right.\\ 
& \left. \left. +\frac{F^{tr}_1}{3}\partial_r R^3-FR\partial_r(F^{tr}_0 R^2) \right\}+ \Phi^2 \left\{ \vphantom{\frac{1}{2}} F^{rr}_0 \partial_rR \partial_r(FR^2) \right.\right.\\
& \left.\left. -F^{rr}_1 R (\partial_r R)^2 - \partial_r \left(F F^{rr}_0 \frac{1}{3}\partial_r R^3\right) \right.\right. \\
& \left.\left. + \partial_r \left(F^{rr}_1 \frac{1}{3} \partial_r R^3\right)\right\}\right] = 0
\end{aligned}
\end{equation}
Now we integrate out the spatial part. By performing the integration over the region between the two boundary points we also get rid off the surface terms as mentioned earlier. This leaves us with purely time-dependent part of the following form 

\begin{equation}\label{time-dep-eq}
\ddot{\Phi} + \epsilon (\mathcal{A}\Phi + \mathcal{B}\dot{\Phi})\dot{\Phi} + \mathcal{C}\Phi + \epsilon \mathcal{D}\Phi^2 = 0
\end{equation}
where the constants $ \mathcal{A},\mathcal{B},\mathcal{C} $ and $ \mathcal{D} $ are given by
\begin{equation}\label{constants}
\begin{aligned}
& \mathcal{A} = \left( \int R^2 F^{tt}_0\right)^{-1} \int \left(\frac{F^{tr}_1}{3}\partial_r R^3-FR\partial_r(F^{tr}_0 R^2) \right)dr\\
& \mathcal{B} = \left( \int R^2 F^{tt}_0\right)^{-1} \int \left(R^3 F^{tt}_1\right)dr \\
& \mathcal{C} = - \left( \int R^2 F^{tt}_0\right)^{-1} \int F^{rr}_0 (\partial_r R)^2dr\\
& \mathcal{D} = \left( \int R^2 F^{tt}_0\right)^{-1} \int \left( F^{rr}_0 \partial_rR \partial_r(FR^2) \right.\\
&\left. -F^{rr}_1 R (\partial_r R)^2\right)dr
\end{aligned}
\end{equation}

Similar to the Newtonian case \cite{Sen2014}, we have found the equation (\ref{time-dep-eq}) to be of the form of general Li{\'e}nard's equation \cite{Strogatz2007,Jordan2007}.

\section{Li{\'e}nard system: fixed points and its stability}\label{Sec:Lienard-system}
Let us first take a look at the $ \epsilon=0 $ case, i.e., the case of linearity stability analysis. From equation (\ref{time-dep-eq}) we see that $ \epsilon = 0 $ gives 
\begin{equation}\label{key}
\ddot{\Phi}+\mathcal{C}\Phi = 0
\end{equation}
which has a solution of the form $ \Phi \propto e^{\pm i\omega_0 t} $, where $ \omega_0 = \sqrt{\mathcal{C}} $. Therefore, if $ \mathcal{C}> 0 $ then the frequency $ \omega_0 $ is real and the solution is oscillatory. On the other hand if $ \mathcal{C}<0 $, the frequency becomes imaginary and the solution becomes hyperbolic and hence the stationary solution becomes unstable. In order to find the sign of $ \mathcal{C} $, it is convenient to write the four velocity components in terms $ u_0 $, where $ u_0 $ is the radial velocity of the infalling matter with respect to the stationary observer \cite{gammie_and_popham,Abramowicz:2016vql}. The four velocity components, i.e., $ v^t $ and $ v^r $ could be given in terms of $ u_0 $ as
\begin{equation}\label{key}
\begin{aligned}
& v^t_0 = \frac{1}{\sqrt{g_{tt}(1-u_0^2)}} \\
& v^r_0 = \frac{u_0}{\sqrt{g_{rr}(1-u_0^2)}}
\end{aligned}
\end{equation}
Now using these transformations, $ F^{tt}_0 $ and $ F^{rr}_0 $ can be written as
\begin{equation}\label{key}
\begin{aligned}
& F^{tt}_0 = \sqrt{\frac{g_{rr}}{g_{tt}}}\frac{u_0}{f_0}\left(\frac{1-u_0^2c_{s0}^2}{1-u_0^2}\right)>0\\
& F^{rr}_0 = \sqrt{\frac{g_{tt}}{g_{rr}}}\frac{u_0}{f_0}\left(\frac{u_0^2-c_{s0}^2}{1-u_0^2}\right)
\end{aligned}
\end{equation}
$ F^{rr}_0 >0 $ for supersonic flow and $ F^{rr}_0<0 $ for subsonic flow, where we have used the fact that $ u_0^2<1,c_{s0}^2<1 $. Thus, one can see from the expression of $ \mathcal{C} $ from equation (\ref{constants}) that for subsonic flow, $ \mathcal{C}>0 $. Therefore, the linear stability analysis suggests that for subsonic flow the stationary accretion solution is stable. This result has been discussed before in literature in the context of linear stability analysis of general relativistic accretion flow, for example, see \cite{Deepika2015,Deepika2017,Shaikh2017,Shaikh2018a,Datta2018}.

Now we turn to the $ \epsilon \neq 0 $ case. In this case the equation (\ref{time-dep-eq}) can recast in the form of the general Li{\'e}nard's equation which is give by \cite{Strogatz2007,Jordan2007}
\begin{equation}\label{Lienerd-eq}
\ddot{\Phi} + \epsilon\mathcal{H}(\Phi,\dot{\Phi})\dot{\Phi}+\mathcal{V}'(\Phi) = 0
\end{equation}
Where `dash' stands for derivative with respect to $ \Phi $. The general Li{\'e}nard's equation is a generalization of the standard Li{\'e}nard's equation (where $ \mathcal{H} $ is function of $ \Phi $ only). The Li{\'e}nard equation is in turn a generalization of well known van der Pol oscillator. It can be interpreted as the equation of motion of a unit mass subject to a nonlinear damping force $ -\epsilon \mathcal{H}(\Phi,\dot{\Phi})\dot{\Phi} $ and a nonlinear restoring force $ -\mathcal{V}'(\Phi) $. Comparing equation (\ref{Lienerd-eq}) with equation (\ref{time-dep-eq}) gives $ \mathcal{H} $ and $ \mathcal{V} $ as
\begin{equation}\label{HandG}
\begin{aligned}
& \mathcal{H}(\Phi,\dot{\Phi}) = \mathcal{A} \Phi + \mathcal{B} \dot{\Phi}\\
& \mathcal{V}(\Phi) = \frac{\mathcal{C}}{2} \Phi^2 + \epsilon \frac{\mathcal{D}}{3}\Phi^3
\end{aligned}
\end{equation}

In order to study the equilibrium points of general Li{\'e}nard system, equation (\ref{Lienerd-eq}) is decomposed into two coupled first order differential equations by introducing a new variable $ \Psi $
\begin{equation}\label{coupled-eq}
\begin{aligned}
& \dot{\Phi} = \Psi \\
& \dot{\Psi} = -\epsilon (\mathcal{A}\Phi + \mathcal{B}\Psi)\Psi - (\mathcal{C}\Phi + \epsilon \mathcal{D}\Phi^2)
\end{aligned}
\end{equation}
The equilibrium points or the `fixed points' of the system is obtained from the condition $ \dot{\Phi} = 0,\dot{\Psi} =0 $. From equation (\ref{coupled-eq}) it is easily found that the fixed points of the system is located at $ (\Phi^\star,\Psi^\star) = (0,0),(-\mathcal{C}/(\epsilon\mathcal{D}),0)$. It is noticed that the fixed points lie on the $ \Psi = 0 $ line. Also, for linear order perturbation one would have only one fixed point at $ (0,0) $ whereas it could be understood that higher order nonlinearity  would result in higher number of fixed points on the $ \Psi =0 $ line.

Now we examine the stability of the fixed points of the system. In order to do that, we perturb the variables $ \Phi$ and $\Psi $ slightly from it's value at the fixed points. In other words, we write 
$ \Phi = \Phi^\star + \delta \Phi $ and $ \Psi = \Psi^\star + \delta \Psi $. Substituting these expressions in equation (\ref{coupled-eq}) and retaining terms upto linear order in $ \delta\Phi $ and $ \delta\Psi $ gives

\begin{equation}\label{lin-pert-lienard-system}
\begin{aligned}
& \delta \dot{\Phi} = \Psi^\star + \delta \Psi = \delta\Psi \\
& \delta \dot{\Psi} = -\mathcal{V}''(\Phi^\star)\delta\Phi-\epsilon \mathcal{H}(\Phi^\star,\Psi^\star)\delta\Psi
\end{aligned}
\end{equation}
where $ \mathcal{V}'' = \mathcal{C}+2\epsilon\mathcal{D}\Phi^\star  $. Now we use trial solution of the form $ \delta\Phi \sim e^{\omega t} $ and $ \delta\Psi \sim e^{\omega t} $. Equation (\ref{lin-pert-lienard-system}) can be written as
\begin{equation}\label{linearized-system}
\begin{bmatrix}
\delta\dot{\Phi} \\
\delta\dot{\Psi}
\end{bmatrix} = \begin{bmatrix}
0 & 1\\
-\mathcal{V}'' & -\epsilon\mathcal{H}
\end{bmatrix}\begin{bmatrix}
\delta\Phi \\
\delta \Psi
\end{bmatrix} = \mathcal{J}\begin{bmatrix}
\delta\Phi \\
\delta \Psi
\end{bmatrix}
\end{equation}
$ \omega $ are given by the eigen velues of the Jacobian $ \mathcal{J} $ defined in the above equation. Therefore, the $ \omega $ are obtained as
\begin{equation}\label{omega}
\begin{aligned}
& \omega_{\pm} = -\epsilon \frac{\mathcal{H}(\Phi^\star,\Psi^\star)}{2}\pm \sqrt{\epsilon^2\left(\frac{ \mathcal{H}(\Phi^\star,\Psi^\star)}{2}\right)^2-\mathcal{V}''(\Phi^\star)}\\
& = \frac{1}{2}\left(\tau \pm \sqrt{\tau^2-4\Delta}\right),\quad \tau = \omega_-+\omega_+,\quad \Delta = \omega_-\omega_+
\end{aligned}
\end{equation}
where $ \tau = -\epsilon \mathcal{H}(\Phi^\star,\Psi^\star)  = -\epsilon \mathcal{A}\Phi^\star$ and $ \Delta = \mathcal{V}'' = \mathcal{C}+\epsilon 2\mathcal{D}\Phi^\star $. We arrive at the system of equation (\ref{linearized-system}) by linearizing the equations (\ref{coupled-eq}), thus equation (\ref{linearized-system}) is the so-called linearized system with Jacobian $ \mathcal{J} $. The stability of the fixed point studied via such linearized system depends on the sign of $ \tau $, $ \Delta $ and $ \tau^2-4\Delta $. To obtain the linearized system we have neglected quadratic terms like $ (\delta \Phi)^2 $ and others. It happens that such linearized system cannot safely tell whether a fixed point is indeed stable or unstable under inclusion of higher order terms for some kind of fixed points. For example, if $ \tau = 0 $ and $ \Delta>0 $ then $ \omega_{\pm} $ are purely imaginary and then the  fixed point for such case is called center-type \cite{Strogatz2007}. For center-type fixed points, the linearized system cannot tell safely whether it will remain so if higher order terms are taken into account (i.e. terms like $ \delta \Phi ^2$). If for example, the $ \omega_\pm $ are real with opposite signs, then the fixed point is called a saddle. Unlike cetre-type fixed point, the linearized system predicts the stability of saddle type fixed points correctly, i.e., it remains saddle type even if higher order terms are included \cite{Andronov1973}. Below we study the stability of the two fixed points of the system (\ref{coupled-eq})  and what implications it has for the accretion flow.

\subsection{Fixed point 1: $ (\Phi^\star,\Psi^\star) = (0,0) $}
For this fixed point $ \tau = 0,\Delta = \mathcal{C} >0$ as it was shown earlier that for subsonic flow $ \mathcal{C}>0 $. Therefore, $ \Delta>0 $ and $ \tau^2-4\Delta <0 $ and hence the $ \omega_\pm $ are purely imaginary and the fixed is centre-type. Such fixed points are surrounded by closed orbits in $ (\Phi,\Psi) $ plane. This is basically identical to the result as obtained in the $ \epsilon =0 $ case, i.e., in the linear perturbation analysis as shown at the beginning of this section. By setting $ \epsilon =0 $, what we obtained was an equation of Harmonic oscillator with frequency $ \sqrt{\mathcal{C}} $. The orbits are closed due to the fact that around this fixed point $ \Phi\ll1,\Psi\ll1 $ and therefore the damping term, as well as the quadratic term of $ \Phi $ in equation (\ref{time-dep-eq}), is negligible and the resulting equation is approximately the harmonic oscillator equation with conserved total energy. However, as mentioned earlier, the linearized system cannot safely predict the stability of center type fixed points. Thus the existence of closed orbit around the center-type fixed point is a consequence of linearization of the system. If we numerically solve the coupled equations (\ref{coupled-eq}) we can understand the actual nature of the phase space trajectories around this fixed point. In fact, it is obvious from the presence of damping term that the system is not conservative and the trajectories will not be closed but rather spiral either outward or inward. From figure \ref{Fig:phase-portrait} it is indeed noticed that the trajectories around the fixed point is a spiral one and the fixed point is actually a stable spiral. However, as the trajectories approach the fixed point at the origin, the damping terms becomes negligible and the trajectories settle into approximately circular orbits.  Let us now see what the linearized system tells us about the stability of the second fixed point.

{
	\centering
	\includegraphics[scale=0.4]{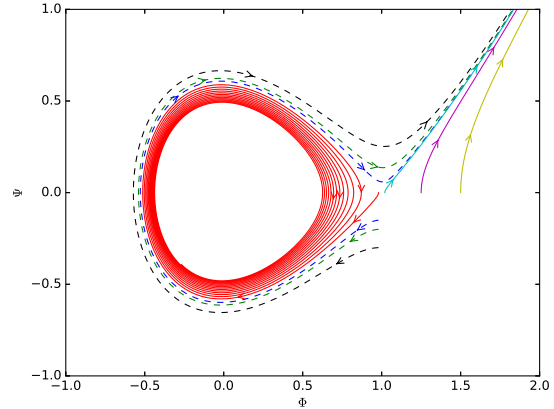}
	\captionof{figure}{\small phase portrait in the $ \Phi-\Psi $ plane. The solid lines (in the $ \phi >1$ region) are for initial value $ (\Phi,\Psi) = (1.02,0),(1.25,0),(1.5,0) $ from left to right. The dashed lines are for initial value $ (\Phi,\Psi) = (0.98,-0.15),(0.98,-0.2),(0.98,-0.3) $ from top to bottom (in the $ \Psi<0 $ region). The spiral in between these lines is for initial value $ (\Phi,\Psi)=(0.98,0) $. We have used $ \mathcal{C}= 1, \mathcal{D} =-1 $ and $ \mathcal{A} = \mathcal{B} = 0.05 $. The nature of the orbit depends on the initial value of $ (\Phi,\Psi) $. In other words, orbits closed to the origin spirals in towards it and hence the perturbation does not grow and the solution becomes stable. On the other hand far from the origin orbits diverges towards infinity and hence perturbation grows with time. Specially orbits on the right side of fixed point $ (-\mathcal{C}/(\epsilon \mathcal{D}),0) $ always diverges. This is due to the fact that for saddle type fixed point, the $ \omega_\pm $ are real and have opposite sign and hence one of the mode will grow exponentially.}\label{Fig:phase-portrait}
}

\subsection{Fixed point 2: $ (\Phi^\star,\Psi^\star) = (-\mathcal{C}/(\epsilon\mathcal{D}),0) $}
The linear stability analysis ($ \epsilon=0 $) predicts one fixed point at the origin of $ (\Phi,\Psi) $ plane which also remains when we include nonlinearity. However, the inclusion of nonlinearity gives rise to a second fixed point at $(\Phi^\star,\Psi^\star) = (-\mathcal{C}/(\epsilon\mathcal{D}),0) $. For this fixed point $ \tau = ({\mathcal{A}\mathcal{C}})/{\mathcal{D}} $ and $ \Delta = -\mathcal{C} <0$. Therefore the $ \omega_\pm $ are real with opposite signs and hence the fixed point is a saddle-type and it will remain so even if higher order terms are included in equation (\ref{linearized-system}). So whatever information we get here will still be valid for the full nonlinear equation (\ref{coupled-eq}). As the $ \omega_\pm $ are real and have opposite sign, one of the modes will grow with time exponentially. And thus the perturbation will grow with time. From the figure \ref{Fig:phase-portrait} it is evident that the trajectories starting on the right side of the fixed point escapes towards infinity. The actual position of the fixed point will, of course, depend on the value and sign of $ \mathcal{C}/\mathcal{D} $. For the purpose of illustration, we have used $ \mathcal{C} = 1 $ and $ \mathcal{D} = -1 $. Therefore the fixed point location is $ (1,0) $. So for $ \Phi>1 $, we get diverging trajectories, on the other hand for $ \Phi<1 $, the trajectories may be spiral if close enough to the origin otherwise diverge. If we change the value and sign of $ \mathcal{C}/\mathcal{D} $, the qualitative features would remain same.

Therefore, it may be concluded that if the initial value of the perturbation is small, i.e, the initial point of the trajectory is close enough to the origin, then the trajectory may spiral towards the origin and settle into nearly circular orbits, making the accretion solution stable. On the other hand, if the initial value of the perturbation is large and the trajectory starts far from the origin it will diverge and hence the stationary accretion solution will be unstable.

The fixed points locations are basically obtained from $ \mathcal{V}'(\Phi^\star) = 0 $. For the present analysis where we include only the lowest order of nonlinearity (i.e., terms of the second order in perturbations) $\mathcal{V}' (\Phi^\star) $ is quadratic in $ \Phi^\star $ whereas for linear perturbation analysis it is linear. As a consequence of this, linear perturbation gives one fixed point and next order perturbation gives two fixed points. If we include higher order perturbations, the number of fixed points will increase which would lie on the $ \Phi $ axis in the $ (\Phi-\Psi) $ plane. In order to say anything about the new fix points one have to carry out the required perturbation analysis which is beyond the scope of the present work \cite{Shaikh2018c}.

\section{Concluding remarks}
We summarize the results as follows: the standard linear stability analysis of global subsonic flows shows that the perturbations are oscillatory in nature and the corresponding accretion flow is stable under such linear perturbation. However, we find that the inclusion of lowest order of nonlinearity in the perturbation scheme affects the results of linear stability analysis considerably. In fact, numerical solution of the resulting equation (Li{\'e}nard equation) suggests that under the influence of the nonlinear term, the previously closed orbits become spirals and the corresponding fixed point becomes stable spiral. Most importantly, a second fixed point appears due to the inclusion of nonlinearity which is a saddle type. We see that perturbations with small enough initial values may become oscillatory with nearly time independent amplitude but if the initial value of perpetuation is not small enough it will diverge exponentially. Thus the inclusion of nonlinearity can make the accretion solution unstable unless the initial value of the perturbation is sufficiently small. 

It is to be mentioned in this regard, that our conclusion about the unstable orbits is based on the study of inviscid flow. However, real fluid is viscous in nature and viscosity often plays a role in opposing the effect of nonlinearity.  In fact, for linear stability analysis, it is noticed that viscosity helps in decaying the amplitude of standing waves \cite{Ray2003}. So the instability arising due to the nonlinearity may be compensated by the viscosity \cite{Stellingwerf1978}. It should also be pointed out the inclusion of higher order nonlinear terms does not make the diverging trajectories decay but it may help to saturate to a value instead of escaping to infinity \cite{Volovik2006,Ray2007}.

We have used the perturbation scheme exclusively for standing wave analysis global subsonic flows. It is to be mentioned that in order that standing wave is formed, the accretion flow must be globally subsonic. This is due to the fact that a supersonic region may develop shock which makes the accretion variables discontinuous at the shock location. 

One can, in principle, extend the analysis to the travelling waves \cite{Petterson1980,Ray2007,Naskar2007,Sarkar2013}. To study travelling waves, one may follow the scheme provided by Petterson {\it et. al.}\cite{Petterson1980} for Newtonian accretion flow. However, for general relativistic accretion, the resulting equations are too complicated to predict analytically anything about the stability even in case of linear perturbation analysis. Using the scheme including nonlinearity is therefore beyond scope of the present work and may be reported elsewhere.

The obvious next step related to the present work would be to use the scheme for axially symmetric flow in the Schwarzschild metric and to even the more general case of axially symmetric flow in the Kerr metric. Such work is in progress and will be reported later.

\bibliographystyle{iopart-num}
\bibliography{reference_arif}

\end{multicols}

\end{document}